# Slidephononics: Tailoring Thermal Transport Properties by van der Waals Sliding

Linfeng Yu[1,2], Chen Shen[1,*], Guangzhao Qin[2,*], Hongbin Zhang[1]

[1]*Institute of Materials Science, Technical University of Darmstadt, Darmstadt 64287, Germany.*

[2]*State Key Laboratory of Advanced Design and Manufacturing for Vehicle Body, College of Mechanical and Vehicle Engineering, Hunan University, Changsha 410082, P. R. China*

**Abstract:**

By interlayer sliding in van der Waals (vdW) materials, the switching electric polarization of ultrathin ferroelectric materials leads to the widely studied slidetronics. In this work, we report that such sliding can further tailor anharmonic effects and hence thermal transport properties due to the changed intrinsic coupling between atomic layers. And we propose an unprecedented concept dubbed as *slidephononics*, where the phonons and associated physical properties can be controlled by varying the intrinsic stacking configurations of slidetronic vdW materials. Based on the *state-of-the-art* first-principles calculations, it is demonstrated that the thermal conductivity of boron nitride (BN) bilayers can be significantly modulated (by up to four times) along the sliding pathways. Detailed analysis reveals that the variation of thermal conductivities can be attributed to the tunable (de-)coupling of the *out-of-plane* acoustic phonon branches with the other phonon modes, which is induced by the interlayer charge transfer. Such strongly modulated thermal conductivity via interlayer sliding in vdW materials paves the way to engineer thermal management materials in emerging vdW electronic devices, which would shed light on future studies of slidephononics.

---

[*] Author to whom all correspondence should be addressed. E-Mail: chenshen@tmm.tu-darmstadt.de (Chen Shen), gzqin@hnu.edu.cn (Guangzhao Qin)

# 1. Introduction

The outstanding physical properties of two-dimensional (2D) materials have attracted intensive attention in the last two decades, as found in quite a few 2D materials such as graphene-like materials[1], carbon monolayer materials[2], transition metal chalcogenides (or halides)[3,4], and layered $MY_2X_4$ materials[5,6]. Such materials, particularly their vdW heterostructures, can be applied to engineering high-performance electronic devices. For such ultrathin devices working in the quantum limit, it is essential to embrace efficient thermal management to maintain their optimal performance and lifetime. The approaches to regulate the thermal conductivity of the thermal management materials can be divided into two categories: firstly crystal structure modification or functionalization, including techniques such as doping[7,8], straining[9–11], hydrogenation[12–14], alloying[15,16], and defects[17,18], and secondly electric field-induced thermal management techniques[19,20]. One significant distinction between these two classes of approaches is that the former class relies mostly on irreversible modification of the intrinsic properties of materials, whereas the second class enables a non-destructive and reversible regulation of heat transport. As the electric fields may intervene with the functioning of the vdW devices, it is appealing to develop a novel approach to control the thermal conductivity of 2D semiconductor materials, which can further promote the application of such materials in micro-nano electronic devices.

Recently, *slidetronics* has made its debut in the 2D materials community[21,22], where the lateral shifts between vdW layers give rise to sign changes in their out-of-plane electric polarization[23–31]. That is, the physical properties can also be tuned by interlayer sliding or different stacking configurations. This is another degree of freedom beyond the twisting in vdW bilayers or multilayers, which lead to intriguing twistronics[32–36]. It was first theoretically demonstrated that the polarization in boron nitride (BN) bilayers can be switched by interlayer sliding[30]. Moreover, the electronic properties of relevant vdW materials can also be tuned for various stacking configurations[21,27,30]. For example, in interlayer-slip ferroelectrics, interlayers lead to unequal upper and lower layers of a 2D material bilayer, leading to a net vertical charge transfer between layers, prompting a reversal of the vertical ferroelectric polarization. The experimentally demonstrated widespread existence of sliding ferroelectricity in van der Waals layered materials[27,30,37–40], where the sliding barrier between layers is several orders of magnitude lower than in conventional ferroelectricity, is anticipated to reduce the energy needed for ferroelectric switching significantly. Additionally, for 2D materials, such as BN and GaX monolayers[41,42], the out-of-plane phonon mode dominates the thermal conductivity and exhibits a potentially large sensitivity to vertical potential fields. So, it's reasonable to wonder whether interlayer sliding may be applied to tailor the phononic thermal transport properties.

In this work, by solving the Boltzmann phonon transport equation based on first-principles calculations, we conduct a comprehensive investigation of the phonon transport properties of BN bilayers in different stacking configurations driven by interlayer sliding. In BN bilayers, we observe that interlayer sliding significantly affects the lattice thermal conductivity without changing the intrinsic bonding between the monolayers. Fundamental insights into sliding-modulated thermal conductivity are obtained by analyzing the phonon anharmonicity, phonon renormalization, and charge transfer. The intrinsic potential field between the BN bilayers is renormalized during sliding as a consequence of the redistribution of the free interlayer charge transfer, which leads to phonon renormalization to modulate thermal conductivity. Dubbing such an effect as slidephononics, our study establishes a novel reversible approach to tailor the thermal transport properties of vdW materials, unlocking a fascinating subject for future research and applications.

## 3. Results and Discussion

### 3.1 Tunable and strong slidephononics

As shown in Fig. 1(a) and (b), the BN bilayers are slid along both the *armchair* and *zigzag* directions to investigate the tunability of the thermal transport properties. For the *armchair* (*zigzag*) interlayer sliding, the upper or lower layer of BN moves along the axial (perpendicular) direction of the bonding axis, respectively. The AA-stacked BN configuration, where the upper boron atom corresponds to the lower nitrogen atom, is considered the pristine configuration. The ratio of the thermal conductivity after slip ($\kappa_s$) to the initial thermal conductivity ($\kappa_i$) is defined as the regulation ratio, $R = \kappa_s/\kappa_i$, which indicates the modulation effect strength of the thermal conductivity. When R > 1, it indicates that the thermal conductivity has an enhanced effect, otherwise it is weakened. Fig. 1(c) and (d) depict sliding configurations with relative local peak and valley thermal conductivities along two sliding paths. High thermal conductivities are found in atomic misalignment (*armchair*) and bonding misalignment (*zigzag*) configurations, respectively, while locally low thermal conductivities occur when atoms (*armchair*) or bonds (*zigzag*) overlap. For instance, the lowest and highest thermal conductivities along the *armchair* path are 460 and 90 W/mK at 300 K, respectively [Fig. 1(e)]. The trend of R is consistent with that of thermal conductivity, and the highest R of ~2 is found in the configuration with 80% slide along the *armchair* direction, while the lowest R of ~0.5 was found at 33% sliding. The wide modulation ratio reveals the high efficiency of the sliding-regulated phonon transport. For sliding along the *armchair* direction, the peaks of thermal conductivities when sliding along the *armchair* direction are found in the case of boron atoms (or nitrogen atoms) located in the middle of the B-N bond, typically at 10 %, 50 %, and 80 % sliding, *i.e.*, atomic misalignment. That is, when the atoms in the upper and lower layers are aligned with 0 % and 33 % sliding, the corresponding thermal conductivities are reduced. Slightly differently, for the *zigzag* sliding, the peaks of thermal conductivities occur when the B-N bonds are mutually misaligned, *i.e.*, 20~30 % and 70~80 % sliding, while the valleys are in the slip configurations of 0% and 50% with bonding overlap. The overall variations of the thermal conductivities along the *armchair* and *zigzag* sliding pathways are shown in Fig. 1(e) and (f), respectively, where significant modulations are clearly visible. It is noted that the sliding barrier along both sliding pathways is marginal, *e.g.*, 35.70 and 17.54 meV per unit cell for the *armchair* and *zigzag* cases, as shown in Fig. 1(g) and (h), respectively. Such an energy barrier can be attributed to the vdW interaction between the layers. The formation of a centrosymmetric configuration is lower in energy than other metastable stacking configurations, and such a tendency is in good agreement with previous first-principles predictions for a two-layered system[21]. Interestingly, the valleys of thermal conductivities exist in both energy valleys and peaks of the migration barrier, regardless of sliding along the *armchair* or *zigzag* direction. This indicates that the sliding configuration with high thermal conductivities can be obtained between the potential valley and the potential peak, *i.e.*, the configuration of atoms (*armchair* direction) and B-N bonding mismatch (*zigzag* direction).

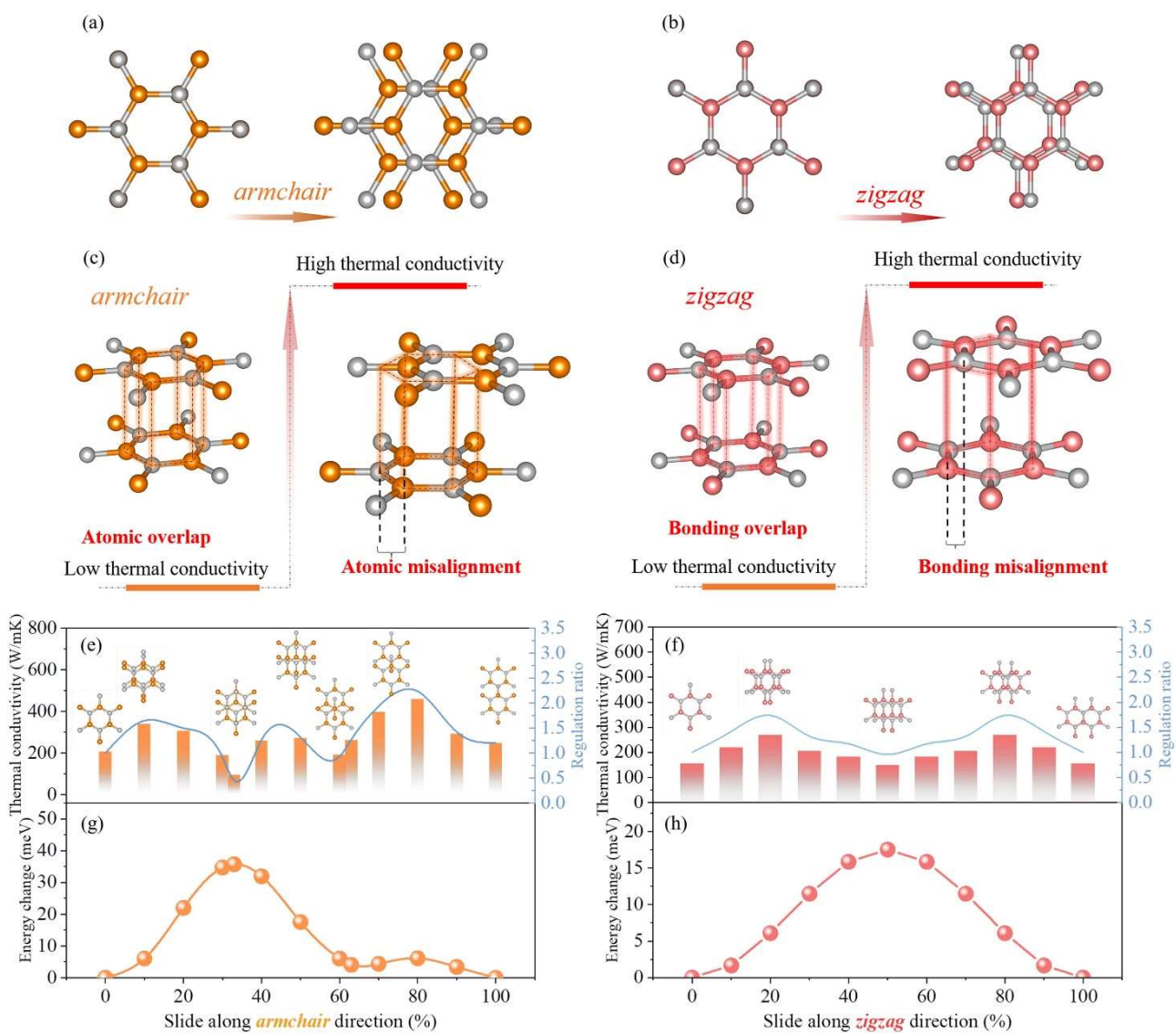

Figure 1. The slidephononics of BN bilayers. Schematic diagram of sliding along the (a) *armchair* and (b) *zigzag* direction. Schematic diagram of slip configurations with high and low thermal conductivities along the (c) *armchair* and (d) *zigzag* direction. Variation of thermal conductivities for slip along the (e) *armchair* and (f) *zigzag* directions. Schematic of the slidephononics of BN bilayers along (g) *armchair* and (h) *zigzag* directions, where the percentages indicate how much the atomic positions slide in fractional coordinates.

## 3.2 Sliding-driven phonon renormalization

Heat transfer in solids is described by phonon interactions, and the phonon dispersion essentially describes the functional relationship between phonon energy and momentum. Traditionally, applying strain breaks the quadratic dispersion relation of the ZA acoustic phonon branch of 2D materials to tune the phonon properties due to strong mechanical stress[9]. Slightly differently, interlayer sliding depends on van der Waals interactions, which are substantially weaker than intrinsic bonding. We have observed that the intrinsic potential field of BN bilayers has redistributed, which may affect the anharmonic interaction between phonons, although the intrinsic bond itself has not changed significantly. Higher-order variables take on greater significance in anharmonic systems because the interaction between phonons is no longer linear. This nonlinearity develops as a result of departures from the ideal harmonic behavior brought on by modifications to the potential energy pattern. The redistribution of the potential well impacts the behavior of the phonons when adjusting the intrinsic energy potential field in a bilayer system. Because phonons are vulnerable to local potential energy, variations in the potential well can affect how they spread and behave. The anharmonic interaction between the phonon and the phonon changes due to the redistribution of the potential well, considering that the intrinsic energy potential field in the bilayer system is dramatically adjusted, as seen in Fig. 1(g) and (h).

To explore the phonon-phonon interaction under different slip potential fields, we will divide the phonon-phonon scattering into three states based on energy conservation: supersaturated ($w_1 + w_2 > w_3$), saturated ($w_1 + w_2 = w_3$), and undersaturated states ($w_1 + w_2 < w_3$). Phonon-phonon scattering may be separated into emission and absorption processes based on energy conservation. In the instance of three phonons, the absorption process results in the merger of two lower-energy phonons into one higher-energy phonon, whereas the emission process results in splitting a higher-energy phonon into two lower-energy phonons. As shown in Fig. 2(a), the supersaturation (and undersaturation) state does not satisfy the energy law in phonon-phonon scattering, where phonons in the two low-frequency regions usually scatter to form higher-frequency (and lower-frequency) phonons that are higher (and lower) than the system phonons energy levels, leading to suppressed phonon-phonon scattering and high thermal conductivity, while phonon-phonon scattering in a saturated state contributes a strong scattering rate. This also still holds true for the process of splitting a phonon into two phonons, *i.e.*, the emission process. Phonons in supersaturated and undersaturated states are less prone to participate in scattering, which enhances the thermal conductivity of the overall system. Whereas in BN bilayers, the sliding potential field can realize phonon state switching in the aforementioned states *via* the (de-)coupling of phonon branches, as discussed below.

To explore the phonon state switching induced by the slidephononics, specific phonon dispersions of BN bilayers as examples are extracted along the *armchair* (0%, 10%, and 33%) and *zigzag* (0%, 20%, and 50%) directions, as shown in Fig. 2(b), (c) and (d). In both the *armchair* and *zigzag* directions, it comes to light that the two out-of-plane vibration branches (ZA and ZA') in the BN bilayers go through analogous transforms: ZA-ZA' decouples with sliding. As shown in Fig. 2(e), these three energy-conserving saturation mechanisms compete with each other in the initial state, intermediate state, and final state of the acoustic phonon branch decoupling, which intuitively reflects the phonon renormalization induced by the decoupling of the acoustic phonon branch. In the BN bilayer, the decoupling process of the acoustic branch of the out-of-plane vibration is further directly captured in Fig. 2(f) and (g), which regulates phonon-phonon scattering and ultimately thermal conductivity.

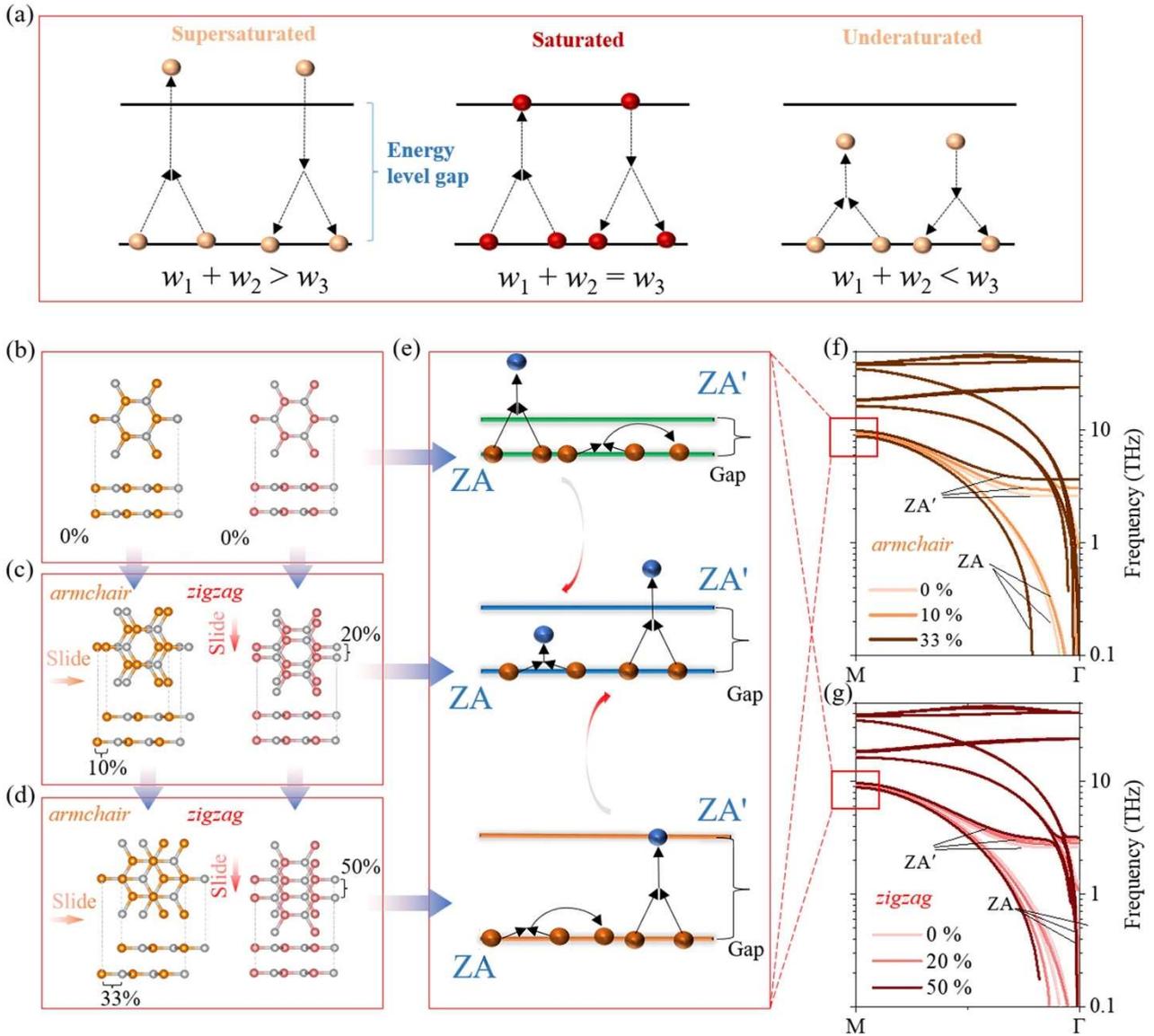

Figure 2. Phonon dispersion and energy conservation. (a) Energy conservation in phonon-phonon scattering. (b) Initial configuration of BN bilayers, which corresponds to the initial sliding configuration. (c) The bilayer configurations, with 10% slip along the *armchair* direction and 20% along the *zigzag* direction, correspond to an intermediate sliding configuration state. (d) The bilayer configurations with 33% slip along the *armchair* direction and 50% along the *zigzag* direction, correspond to the sliding configuration's final state. (e) Schematic diagram of slip-regulated phonon-phonon scattering. The phonon dispersion along (f) the *armchair* direction and along (g) the *zigzag* direction.

**3.3 Competing phonon scattering and scattering channels**

We further monitor the real-time phonon properties of different slide configurations due to the phonon renormalization induced by the transition of the saturation regime. As shown in Fig. 3(a) and (d), the difference in thermal conductivity during the sliding process is mainly contributed by the low-frequency phonons of 0–15 THz, corresponding to the ZA, ZA' branches. Interestingly, the change of phonon group velocity is not obvious, even the group velocity of 10% (20 %) sliding lattice along the *armchair* (*zigzag*) direction is slightly lower than that of 0% and 33% (50%) sliding lattice as shown in Fig. 3(b) and (e), but it has higher thermal conductivity. The interlayer sliding mainly affects the interlayer van der Waals force, and has little effect on the strong covalent bond in the layer, and does not change the chemical composition of the system. The phonon group velocity reflecting the harmonic nature is mainly determined by the strength of chemical bonds and atomic mass[43]. Thus, the group velocity exhibits a weak modulating effect on the thermal conductivity. However, the interlayer sliding breaks the out-of-plane symmetry, leading to an anharmonic change in the potential energy. Hence, the phonon relaxation time, representing an anharmonic feature, will change significantly due to phonon renormalization. As shown in Fig. 3(c) and (f), the relaxation time of the 10% (20%) sliding lattice along the *armchair* (*zigzag*) direction is significantly larger than that of the 0% and 33% (50%) sliding lattice.

Large relaxation times indicate weak phonon-phonon scattering dominated by moderate ZA-ZA' bandgap [Fig. 2(e)]. To further reveal the relationship between phonon branch coupling and phonon-phonon scattering, we decompose different phonon-phonon scattering channels based on the energy conservation process. In the low-frequency region in the initial state, the phonon-phonon scattering of BN bilayers is mainly contributed by ZA phonons, *i.e.*, ZA+ZA→ZA and ZA→ZA+ZA, as shown in Fig. 4(a) and (d). Taking the high symmetry point M as an example, a bandgap of (0.059) 0.199 THz is found between the ZA and ZA' branches when sliding along the *armchair* (*zigzag*) direction. Such a narrow phonon band gap is insufficient to satisfy the fusion of two ZA phonons into a relatively high-frequency ZA' phonon due to the energy conservation of "oversaturation", as shown in Fig. 2(e). When sliding into the intermediate state, the narrow phonon channel between ZA and ZA' at point

M is slightly opened to 0.359 (*armchair*) and 0.697 (*zigzag*) THz. The slightly moderately wide bandgap harmonizes the phonon-phonon scattering of ZA and ZA', suppressing the strong triple ZA phonon-scattering channel as shown in Fig. 4(b) and (e). At this time, the phonon channels participated by ZA and ZA' are undersaturated, and they strongly do not satisfy energy conservation and contribute a large thermal conductivity, as shown in Fig. 2(e), *i.e.*, the saturated state is transformed into supersaturated and undersaturated states. When sliding further to the final state, the bandgap is opened to the maximum, *i.e.*, 1.059 (0.850) THz for the *armchair* (*zigzag*) direction, implying the maximum decoupling of the acoustic phonon branch. The scattering channels participated by ZA and ZA' phonons both reach the strongest saturation state. At this time, the wide band gap allows ZA to cross the band gap and scatter with higher-frequency ZA' phonons. The transition of the scattering channel reveals a significant regulatory effect of slip modulation on phonon-phonon scattering through the decoupling of the acoustic phonon branch.

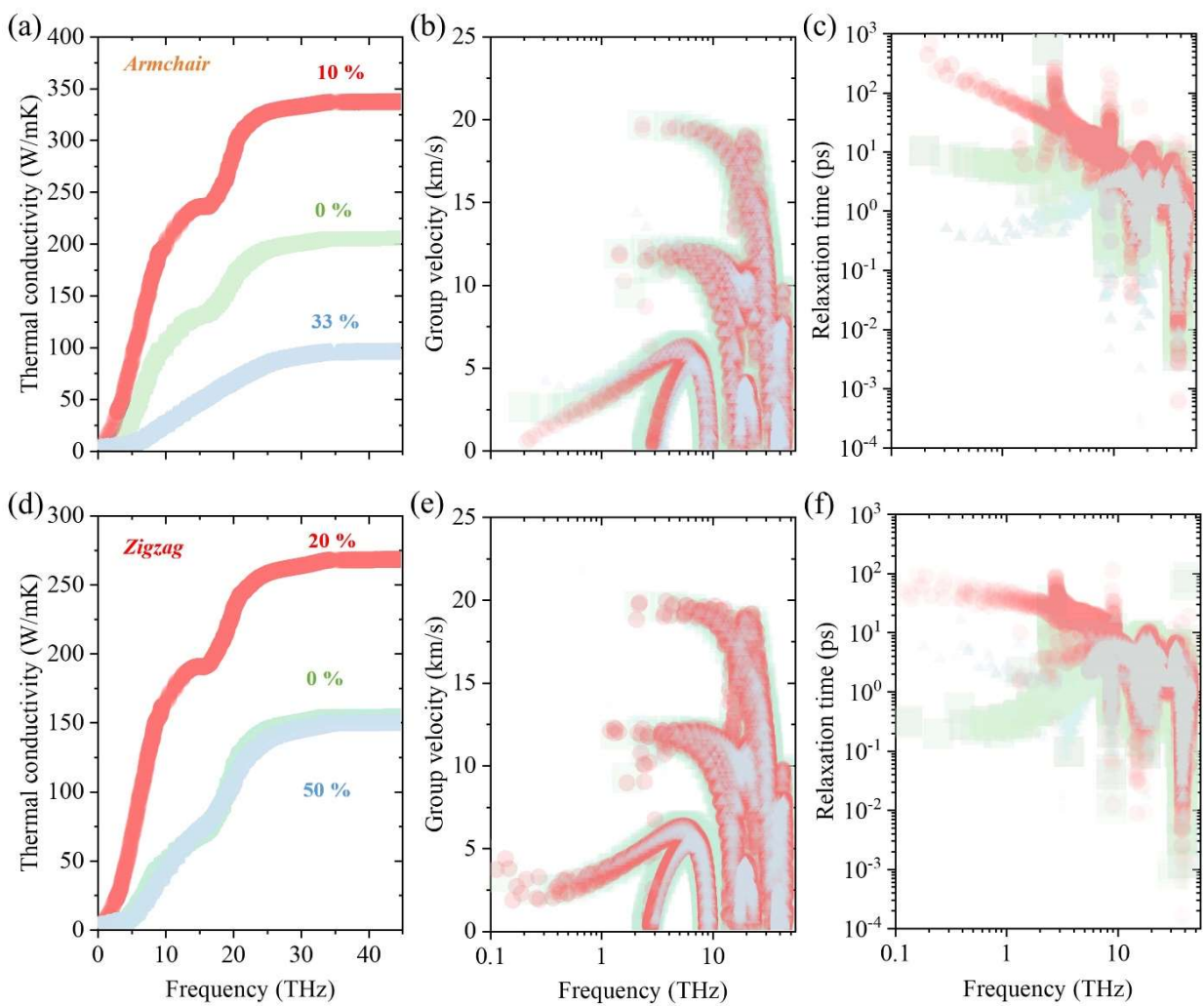

Figure 3. Mode-level phonon properties. (a) Cumulative thermal conductivity, (b) group velocity, and (c) relaxation time along the *armchair* direction. (d) Cumulative thermal conductivity, (e) group velocity, and (f) relaxation time along the *zigzag* direction.

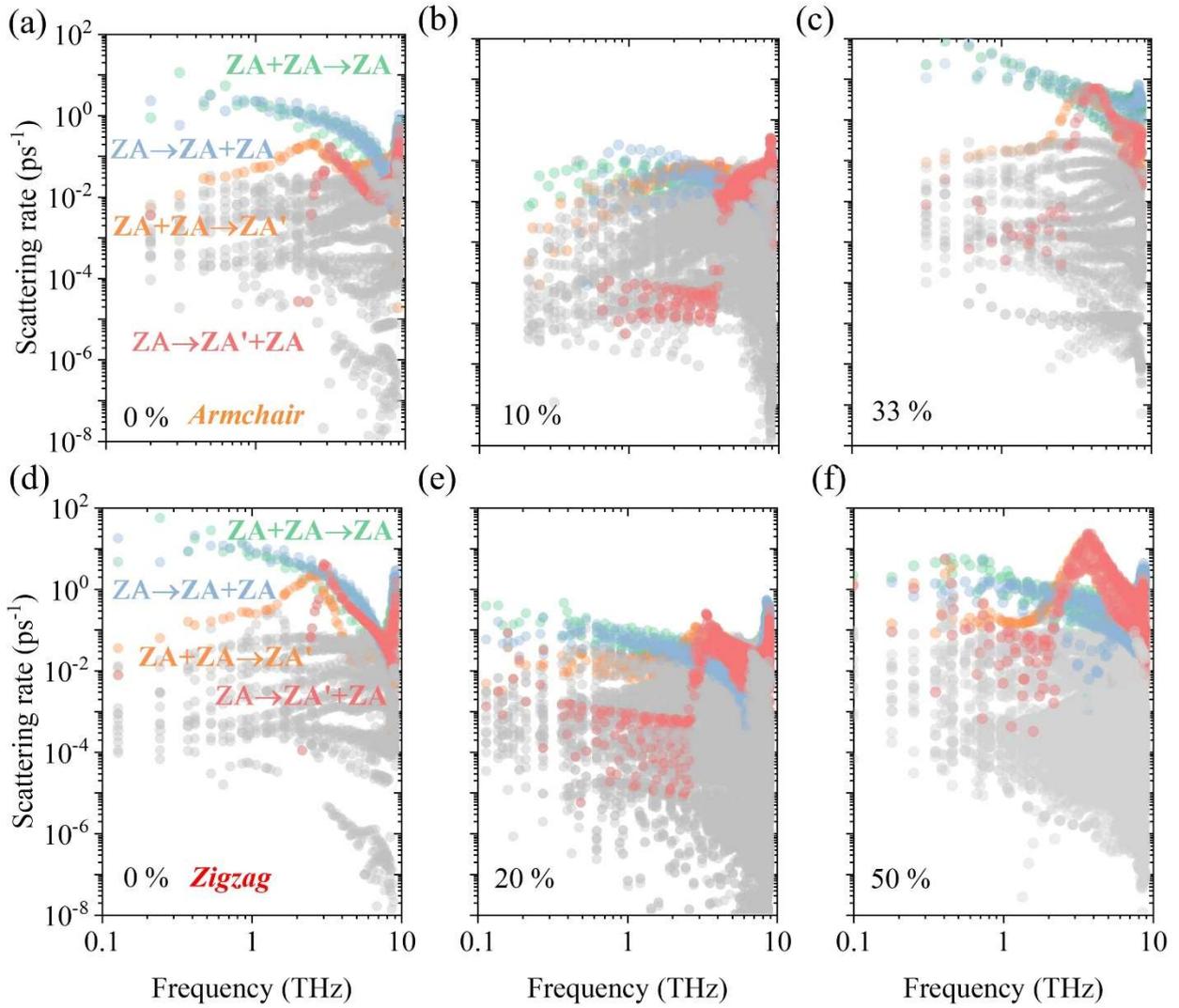

Figure 4. Transition of the slip-regulated scattering channel. Mode level scattering channel for slip (a) 0%, (b) 10% and (c) 33% along *armchair* direction. Mode level scattering channel for slip (d) 0%, (e) 20% and (f) 50% along *zigzag* direction.

**3.4 Sliding-modulated interlayer charge interaction**

Phonon transport is essentially achieved through charge interactions, thus fundamental insights into the thermal transport of interlayer slip can be obtained through interlayer charge interaction transitions. Intrinsic bonding is almost unaffected during sliding, resulting in little change in the group velocity dominated by harmonic properties. Relative to the fixed bonding electrons, the non-bonding charges are free around the atom and are significantly affected by the slip order. Fig. 5(a) and (b) demonstrate the in-plane charge variation along the vacuum layer. Significant charge transitions originate from interlayer charges, ~10Å, while relatively few transitions are inherent to bonding electrons between layers, revealing a strong tunable effect of slip on interlayer charge. In thermal vibrations, atomic anharmonicity originates from the nonlinear Coulomb repulsion between free electrons and bonded electrons.[44,45] As shown in Fig. 5(c), the three-dimensional charge differential density reveals that interlayer charge transfer plays a dominant role. As revealed by the electron localization function in Fig. 5(d), slidephononics are generally non-intrusive, with bonding electrons within the layer barely affected. Although the intrinsic bonding is barely changed, the slip redistribution of interlayer charges enhances the well anharmonicity of atomic vibrations, as revealed in Fig. 1(e–h). The force of the nonlinear force interaction is caused by the van der Waals interaction, which is equivalent to applying a vdW force field in the out-of-plane direction. When the vdW force field is regulated by the slip order, the interlayer interaction charges will be redistributed to cause fluctuations in the potential energy surface of the lattice, resulting in phonon renormalization. In BN bilayers, significant fluctuations are found along the acoustic phonon branch of out-of-plane vibrations. Therefore, the critical feature of slidephononics is that it does not change or destroy the intrinsic bonding of materials, reflecting the non-destructive properties of phonon transport regulation.

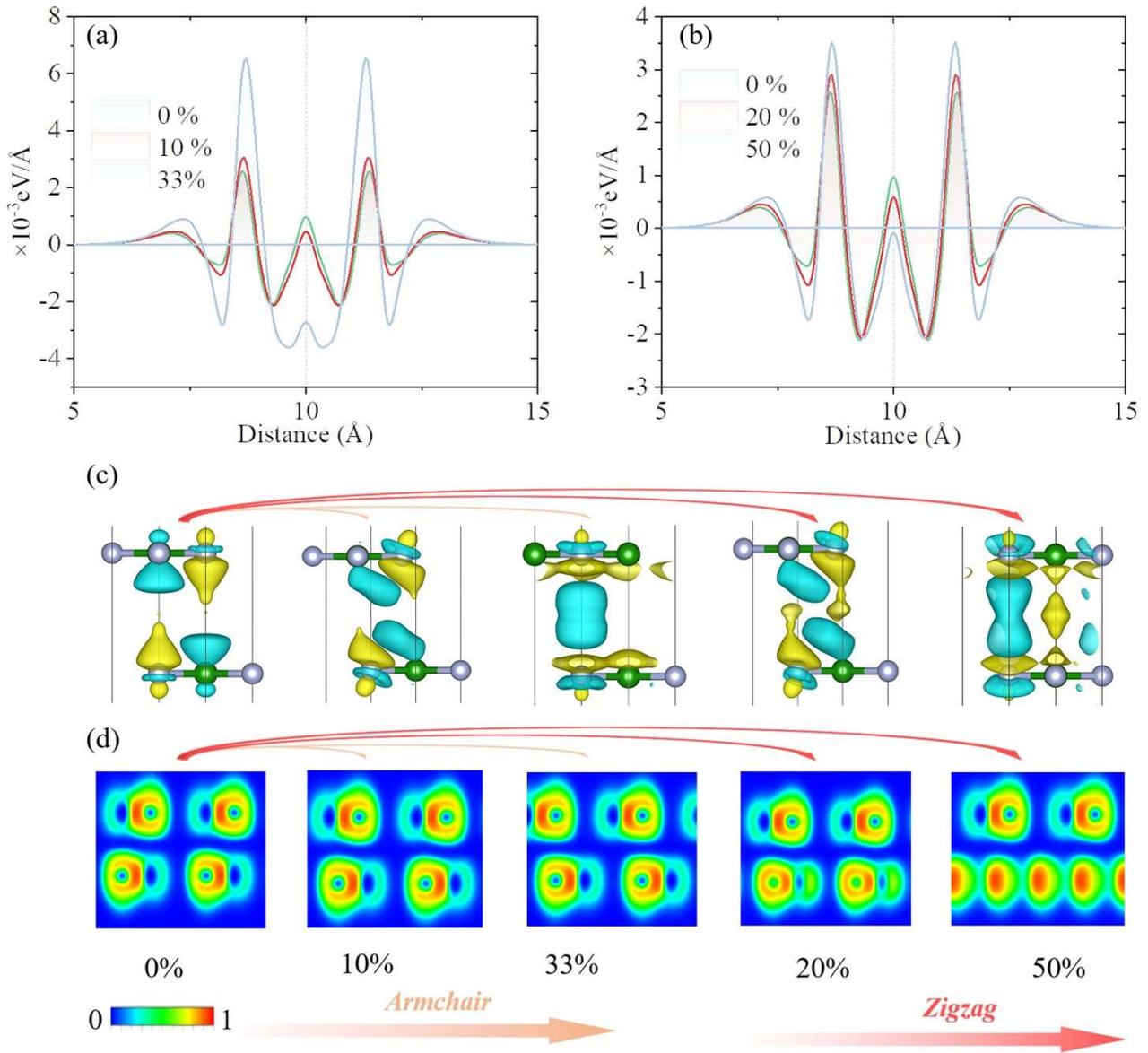

Figure 5. Charge interaction renormalization. (a) The transition of the planar charge density along *armchair* direction. (b) The transition of the planar charge density along *zigzag* direction. (c) Three-dimensional charge density transfer image, where charge transfer mainly occurs between layers at surface $1 \times 10^{-4}$. (d) Electron localization function, where the intrinsic bonds are barely affected. It should be mentioned that the ELF in the 50% slip configuration has a larger mismatch in the same section, but the intrinsic bonding is hardly affected.

## 4. Conclusion

In summary, we propose the concept of slidephononics, which is significant as it provides a novel and controllable approach to tuning the phonon properties of materials. Based on first-principles calculations, sliding phononics demonstrates a wide modulation ratio of thermal conductivity in BN bilayers. The deep fundamental insight is that sliding the layers renormalizes the interlayer charge and causes the phonon renormalization. It achieves the tuning of thermal conductivity by decoupling the acoustic phonon branch, thereby transforming the phonon-phonon scattering mechanism competitively. Compared with traditional methods such as doping or applying an external forcing field, sliding phononics does not change the chemical composition of the material structure but instead by sliding material layers. This regulation is highly controllable and reversible, making it a promising technique for applications in fields such as materials science and electronics. This technique could be used to design devices with improved stability and durability and to study novel quantum effects and spin-phonon phenomena.

## 2. Computational Methodology

The basic theory to perform the calculation is the density functional theory, with the projected augmented wave (PAW) pseudopotential, which is implemented in the Vienna ab initio simulation package (VASP)[46], and the exchange-correlation functional was approximated by the Perdew-Burke-Ernzerh of generalized gradient approximation (GGA-PBE). The kinetic energy cutoff was set as 400eV. In the self-consistent calculation, the k-mesh was set to be 24 × 24 × 1 in Monkhorst-Pack [47] to sample the Brillouin Zone (BZ), and the convergence energy threshold is $10^{-8}$ eV. Van der Waals interactions are corrected via the optB86b parameter[48]. To perform interatomic force constant calculation, a 6 × 6 × 1 supercell with 72 atoms was generated, and a Monkhorst-Pack k-mesh of 4 × 4 × 1 was used to sample the Brillouin Zone. In order to reduce computational resources, it is necessary to ignore the interactions between atoms at a certain large distance, so the cutoff radius ($r_{cutoff}$) is introduced. The cutoff interactions was chose up to 13$^{th}$ based on the convergence test $\kappa$ v.s. $r_{cutoff}$.[49] The $\kappa$ is calculated by solving the linearized BTE[50].

**ACKNOWLEDGEMENTS**

The Lichtenberg high performance computer of the TU Darmstadt is gratefully acknowledged for the computational resources where the calculations were conducted for this project. Y.L. and G.Q. are supported by the National Natural Science Foundation of China (Grant No. 52006057) and the Fundamental Research Funds for the Central Universities (Grant No. 531119200237).

**DECLARATION OF INTERESTS**

The authors declare no competing interests.

**References**

1. Xu, M., Liang, T., Shi, M. & Chen, H. Graphene-Like Two-Dimensional Materials. *Chem. Rev.* 33 (2013).
2. Fan, Q. *et al.* Biphenylene network: A nonbenzenoid carbon allotrope. *Science* **372**, 852–856 (2021).
3. Du, Z. *et al.* Conversion of non-van der Waals solids to 2D transition-metal chalcogenides. *Nature* **577**, 492–496 (2020).
4. Pan, J. *et al.* Auxetic two-dimensional transition metal selenides and halides. *npj Comput Mater* **6**, 154 (2020).
5. Wang, Q. *et al.* Efficient Ohmic contacts and built-in atomic sublayer protection in MoSi2N4 and WSi2N4 monolayers. *npj 2D Mater Appl* **5**, 1–9 (2021).
6. Hong, Y.-L. *et al.* Chemical vapor deposition of layered two-dimensional MoSi2N4 materials. *Science* **369**, 670–674 (2020).
7. Li, Y. *et al.* Ultralow thermal conductivity of BaAg2SnSe4 and the effect of doping by Ga and In. *Materials Today Physics* **9**, 100098 (2019).
8. Liu, W. *et al.* Ag doping induced abnormal lattice thermal conductivity in Cu 2 Se. *Journal of Materials Chemistry C* **6**, 13225–13231 (2018).
9. Pereira, L. F. C. & Donadio, D. Divergence of the thermal conductivity in uniaxially strained graphene. *Physical Review B* **87**, 125424 (2013).
10. Parrish, K. D., Jain, A., Larkin, J. M., Saidi, W. A. & McGaughey, A. J. Origins of thermal conductivity changes in strained crystals. *Physical Review B* **90**, 235201 (2014).
11. Lindsay, L. *et al.* Phonon thermal transport in strained and unstrained graphene from first principles. *Phys. Rev. B* **89**, 155426 (2014).

12. Son, J. *et al.* Hydrogenated monolayer graphene with reversible and tunable wide band gap and its field-effect transistor. *Nat Commun* **7**, 13261 (2016).
13. Wu, X. *et al.* Hydrogenation of penta-graphene leads to unexpected large improvement in thermal conductivity. *Nano letters* **16**, 3925–3935 (2016).
14. Liu, Z., Wu, X. & Luo, T. The impact of hydrogenation on the thermal transport of silicene. *2D Mater.* **4**, 025002 (2017).
15. Bai, G. *et al.* Tailoring interface structure and enhancing thermal conductivity of Cu/diamond composites by alloying boron to the Cu matrix. *Materials Characterization* **152**, 265–275 (2019).
16. Wang, H. *et al.* The exceptionally high thermal conductivity after 'alloying' two-dimensional Gallium Nitride (GaN) and Aluminum Nitride (AlN). *Nanotechnology* (2020) doi:10.1088/1361-6528/abd20c.
17. Hashimoto, A., Suenaga, K., Gloter, A., Urita, K. & Iijima, S. Direct evidence for atomic defects in graphene layers. *Nature* **430**, 870–873 (2004).
18. Bradac, C., Gao, W., Forneris, J., Trusheim, M. E. & Aharonovich, I. Quantum nanophotonics with group IV defects in diamond. *Nat Commun* **10**, 5625 (2019).
19. Deng, S. *et al.* Electric-field-induced modulation of thermal conductivity in poly (vinylidene fluoride). *Nano Energy* **82**, 105749 (2021).
20. Qin, G., Qin, Z., Yue, S.-Y., Yan, Q.-B. & Hu, M. External electric field driving the ultra-low thermal conductivity of silicene. *Nanoscale* **9**, 7227–7234 (2017).
21. Li, L. & Wu, M. Binary Compound Bilayer and Multilayer with Vertical Polarizations: Two-Dimensional Ferroelectrics, Multiferroics, and Nanogenerators. *ACS Nano* **11**, 6382–6388 (2017).
22. Wu, M. & Zeng, X. C. Bismuth Oxychalcogenides: A New Class of Ferroelectric/Ferroelastic Materials with Ultra High Mobility. *Nano Lett.* **17**, 6309–6314 (2017).
23. Wilson, N. P., Yao, W., Shan, J. & Xu, X. Excitons and emergent quantum phenomena in stacked 2D semiconductors. *Nature* **599**, 383–392 (2021).
24. Miao, L.-P. *et al.* Direct observation of geometric and sliding ferroelectricity in an amphidynamic crystal. *Nat. Mater.* **21**, 1158–1164 (2022).
25. Zhang, D., Schoenherr, P., Sharma, P. & Seidel, J. Ferroelectric order in van der Waals layered materials. *Nat Rev Mater* **8**, 25–40 (2023).
26. Rogée, L. *et al.* Ferroelectricity in untwisted heterobilayers of transition metal dichalcogenides. *Science* **376**, 973–978 (2022).
27. Vizner Stern, M. *et al.* Interfacial ferroelectricity by van der Waals sliding. *Science* **372**, 1462–1466 (2021).
28. Wang, X. *et al.* Interfacial ferroelectricity in rhombohedral-stacked bilayer transition metal dichalcogenides. *Nat. Nanotechnol.* **17**, 367–371 (2022).
29. Higashitarumizu, N. *et al.* Purely in-plane ferroelectricity in monolayer SnS at room temperature. *Nat Commun* **11**, 2428 (2020).
30. Yasuda, K., Wang, X., Watanabe, K., Taniguchi, T. & Jarillo-Herrero, P. Stacking-engineered ferroelectricity in bilayer boron nitride. *Science* **372**, 1458–1462 (2021).
31. Wang, C., You, L., Cobden, D. & Wang, J. Towards two-dimensional van der Waals ferroelectrics. *Nat. Mater.* **22**, 542–552 (2023).

32. Cao, Y. *et al.* Correlated insulator behaviour at half-filling in magic-angle graphene superlattices. *Nature* **556**, 80–84 (2018).
33. Kerelsky, A. *et al.* Maximized electron interactions at the magic angle in twisted bilayer graphene. *Nature* **572**, 95–100 (2019).
34. Park, J. M., Cao, Y., Watanabe, K., Taniguchi, T. & Jarillo-Herrero, P. Tunable strongly coupled superconductivity in magic-angle twisted trilayer graphene. *Nature* **590**, 249–255 (2021).
35. Cao, Y. *et al.* Unconventional superconductivity in magic-angle graphene superlattices. *Nature* **556**, 43–50 (2018).
36. Cheng, Y. *et al.* Magic angle in thermal conductivity of twisted bilayer graphene. *Materials Today Physics* **35**, 101093 (2023).
37. Fei, Z. *et al.* Ferroelectric switching of a two-dimensional metal. *Nature* **560**, 336–339 (2018).
38. Bao, Y. *et al.* Gate-Tunable In-Plane Ferroelectricity in Few-Layer SnS. *Nano Lett.* **19**, 5109–5117 (2019).
39. Chang, K. *et al.* Discovery of robust in-plane ferroelectricity in atomic-thick SnTe. *Science* **353**, 274–278 (2016).
40. Wu, M. & Li, J. Sliding ferroelectricity in 2D van der Waals materials: Related physics and future opportunities. *Proceedings of the National Academy of Sciences* **118**, e2115703118 (2021).
41. Yu, L.-F. *et al.* Realizing ultra-low thermal conductivity by strong synergy of asymmetric geometry and electronic structure in boron nitride and arsenide. *Rare Met.* **42**, 210–221 (2023).
42. Shen, C. *et al.* Two-dimensional buckling structure induces the ultra-low thermal conductivity: a comparative study of the group GaX (X = N, P, As). *J. Mater. Chem. C* **10**, 1436–1444 (2022).
43. Yu, L. *et al.* Abnormal enhancement of thermal conductivity by planar structure: A comparative study of graphene-like materials. *International Journal of Thermal Sciences* **174**, 107438 (2022).
44. Skoug, E. J. & Morelli, D. T. Role of Lone-Pair Electrons in Producing Minimum Thermal Conductivity in Nitrogen-Group Chalcogenide Compounds. *Phys. Rev. Lett.* **107**, 235901 (2011).
45. Morelli, D. T., Jovovic, V. & Heremans, J. P. Intrinsically Minimal Thermal Conductivity in Cubic I − V − VI 2 Semiconductors. *Phys. Rev. Lett.* **101**, 035901 (2008).
46. Kresse, G. & Hafner, J. *Ab initio* molecular-dynamics simulation of the liquid-metal–amorphous-semiconductor transition in germanium. *Phys. Rev. B* **49**, 14251–14269 (1994).
47. Monkhorst, H. J. & Pack, J. D. Special points for Brillouin-zone integrations. *Phys. Rev. B* **13**, 5188–5192 (1976).
48. Klimeš, J., Bowler, D. R. & Michaelides, A. Van der Waals density functionals applied to solids. *Phys. Rev. B* **83**, 195131 (2011).
49. Qin, G. & Hu, M. Accelerating evaluation of converged lattice thermal conductivity. *npj Comput Mater* **4**, 3 (2018).
50. Li, W., Carrete, J., A. Katcho, N. & Mingo, N. ShengBTE: A solver of the Boltzmann transport equation for phonons. *Computer Physics Communications* **185**, 1747–1758 (2014).